\documentclass[twocolumn]{aastex62}
\usepackage{apjfonts}
\usepackage{amsmath}
\newcommand{\Hi}{\hbox{H{\sc i}}}

\begin{document}

\title{The role of \Hi~ in regulating size growth of local galaxies}
\shortauthors{Pan et al.}
\shorttitle{The H{\sc i}-to-optical disk size ratio of SFGs}
\author{Zhizheng Pan}
\email{panzz@pmo.ac.cn}
\affiliation{Purple Mountain Observatory, Chinese Academy of Sciences, 8 Yuan Hua Road, Nanjing, Jiangsu 210033, China}

\author{Jing Wang}
\affiliation{Kavli Institute for Astronomy and Astrophysics, Peking University, Yi He Yuan Lu 5, Hai Dian District, Beijing 100871, China}

\author{Xianzhong Zheng}
\affiliation{Purple Mountain Observatory, Chinese Academy of Sciences, 8 Yuan Hua Road, Nanjing, Jiangsu 210033, China}

\author{Xu Kong}
\affiliation{CAS Key Laboratory for Research in Galaxies and Cosmology, Department of Astronomy, \\
University of Science and Technology of China, Hefei, Anhui 230026, China}

\begin{abstract}
We study the role of atomic hydrogen (\Hi) in regulating size growth of local galaxies. The size of a galaxy, $D_{\rm r,~25}$, is characterized by the diameter at which the $r-$band surface brightness reaches $\mu_{\rm r}=25.0~\rm mag~arcsec^{-2}$. We find that the positions of galaxies in the size ($D_{\rm r,~25}$)$-$stellar mass ($M_{\ast}$) plane strongly depend on their \Hi-to-stellar mass ratio ($M_{\rm HI}/M_{\ast}$). In the \Hi-rich regime, galaxies that are more rich in \Hi~tend to have larger sizes. Such a trend is not seen in the \Hi-poor regime, suggesting that size growth is barely affected by the \Hi~content when it has declined to a sufficiently low level. An investigation of the relations between size, $M_{\rm HI}/M_{\ast}$ and star formation rate (SFR) suggests that size is more intrinsically linked with $M_{\rm HI}/M_{\ast}$, rather than SFR. We further examine the H{\sc i}-to-stellar disk size ratio ($D_{\rm HI}/D_{\rm r,~25}$) of galaxies and find that at log($M_{\rm HI}/M_{\ast})>-0.7$, $D_{\rm HI}/D_{\rm r,~25}$ is weakly correlated with $M_{\ast}$. These findings support a picture in which the \Hi-rich galaxies live in an inside-out disk growing phase regulated by gas accretion and star formation. The angular momentum of the accreted materials is probably the key parameter in shaping the size of an \Hi-rich galaxy.

\end{abstract}

\section{Introduction}\label{Sec1}
The sizes of galaxies play a critical role in our understanding of how they form and evolve. Observationally, the absence of a clear border of a galaxy makes the measuring of its extent a non-trivial task. In the literature, there are two popular approaches for characterizing the size of a galaxy. The first, which is also the most popular approach, is defining the size of a galaxy as the radial distance that enclosing half of its light, i.e., the effective radius $R_{\rm e}$ \citep{de Vaucouleurs 1948}. A second commonly used approach is defining galaxy size at the radial location of a given isophote (for example $R_{\rm 25}$, the radius at which the surface brightness reaches $\mu=25.0~\rm mag~arcsec^{-2}$). Each approach has its superiority and shortcoming. As already known, $R_{\rm e}$ is barely affected by the depth of image but is quite sensitive to the light profile of a galaxy, making it incapable of describing the global extension of a galaxy \citep{Graham 2019}. The second approach performs better in describing the global size of a galaxy but requires the image to reach a certain depth.

Observational studies have demonstrated that the sizes of galaxies evolve across cosmic time \citep[e.g.,][]{vanderwel 2014}. On the one hand, mergers are believed to significantly contribute to this evolution, especially at the high mass end. Simulations suggest that the size of the emerged galaxy is strongly dependent on the physical conditions (stellar mass, gas mass, etc) of progenitor galaxies before merging \citep{Hopkins 2009}. On the other hand, in a merger-free case, theoretical studies predict that star-forming galaxies (SFGs) grow from the inside out via gas accretion and star formation \citep[e.g.,][]{Pichon 2011}. This inside-out galaxy formation picture is supported by a growing body of observational studies \citep{Wang 2011,Dale 2016,Frankel 2019,Chen 2020}.

The second scenario points out a potential link between size growth and the cold gas content of a galaxy. Neutral atomic hydrogen (\Hi) gas is the raw material from which molecular gas and then star formation form. Studying the scaling relations between galaxy properties and their \Hi~content thus provides important insights on understanding galaxy formation. Based on the data provided by the GALEX Arecibo SDSS Survey (GASS), \citet{Catinella 2010} showed that the concentration of galaxies is only weakly correlated with their \Hi~content. Subsequent studies found that \Hi-rich galaxies have bluer outer disk  than the \Hi-normal ones, implying continuous growth of the outer disks supplied by rich \Hi~gas reservoirs \citep{Wang 2011,Huang 2014,  Kauffmann 2015,Yildiz 2017}. Recently, \citet{Chen 2020} study the relations between the properties of bulge/disk component and the \Hi~content of galaxies. They found that the color of disk is bluer in \Hi~rich galaxies, whereas a similar trend is not found for the bulge component. This finding suggests that \Hi~gas is closely related to the formation of disks but not necessarily fuel the star formation of bulges in an efficient way.

\begin{figure*}
\centering
\includegraphics[width=140mm,angle=0]{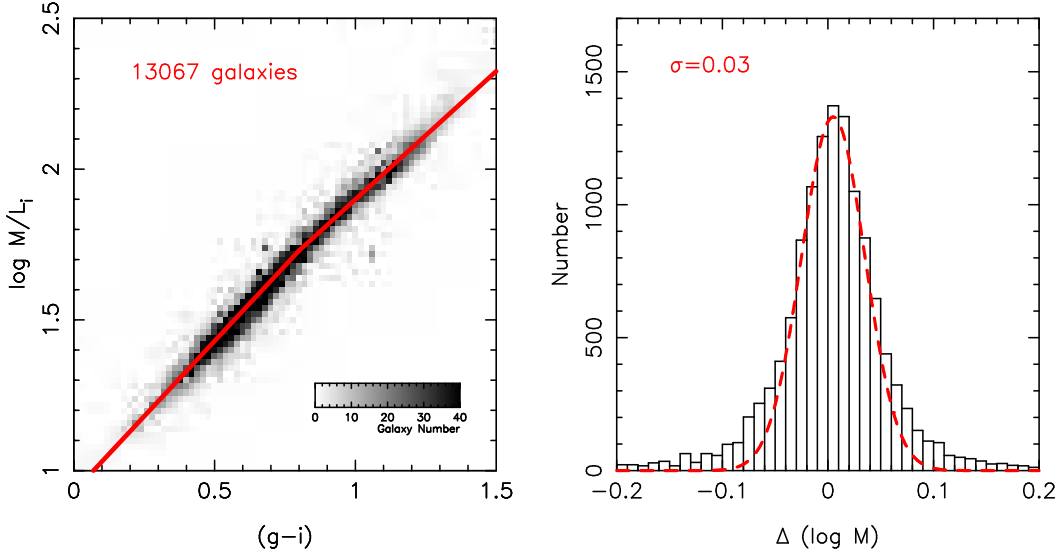}
\caption{Left: the correlation between mass-to-light ratio $M/L_i$ and the SDSS $g-i$ color index. We fit the relation with a broken linear relation, as shown in the red line. Right: the $\Delta~M$ distribution of the spectroscopic-z sample, where $\Delta~M$ is the stellar mass difference between that given by the JHU/MPA catalog and that estimated using the $M/L_i$ vs.($g-i$) relation. The red dashed line shows a Gaussian function with $\sigma=0.03$.  }\label{fig1}
\end{figure*}

\begin{figure*}
\centering
\includegraphics[width=140mm,angle=0]{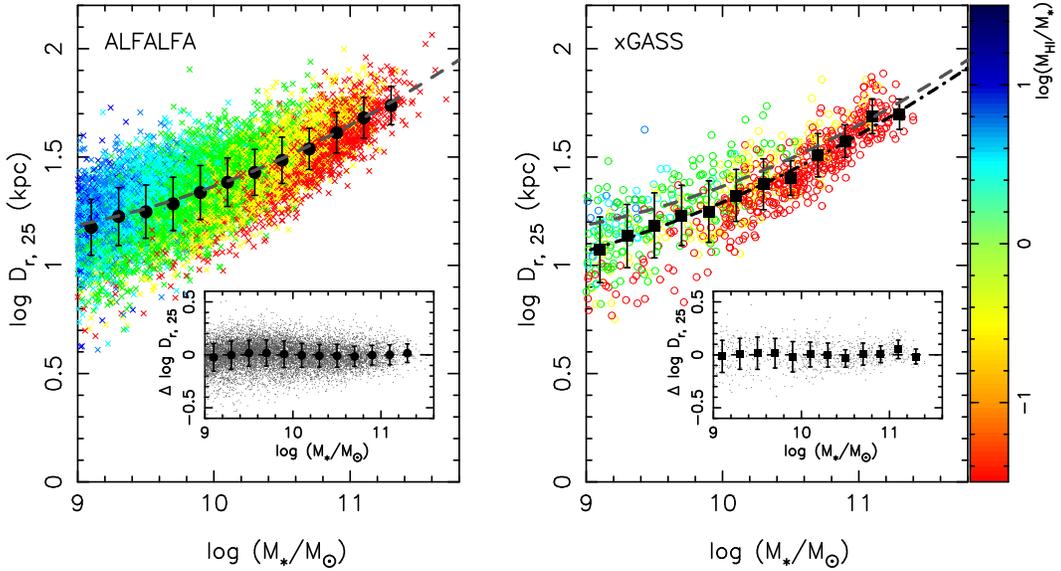}
\caption{Left: the size$-$mass relation of the ALFALFA sample. The large symbols represent the median log$(D_{\rm r,~25})$ in that $M_{\ast}$ bin, with a bin size of $\Delta$log$M_{\ast}=0.2$. Error bars represent the 1$\sigma$ standard deviation. We use a second-order polynomial to fit this relation, and the best-fit result is shown in the gray dashed line. Small symbols are color coded by log$(M_{\rm HI}/M_{\ast})$. The small panel shows the residuals of the fit against stellar mass. Right: the size$-$mass relation of the xGASS sample, with its best fit result shown in black dash-dotted line. }\label{fig2}
\end{figure*}

Given the fact that \Hi~content is closely related to the formation of disk component of galaxies, it is worthy to further investigate at what condition the \Hi~gas could efficiently impact on size growth. In this paper, we aim to study how size growth is regulated by the \Hi~gas of galaxies based on the data from the extended GASS (xGASS, \citealt{Catinella 2018} ) and Arecibo Legacy Fast ALFA (ALFALFA) survey. Throughout this paper, we adopt a concordance $\Lambda$CDM cosmology with $\Omega_{\rm m}=0.3$, $\Omega_{\rm \Lambda}=0.7$, $H_{\rm 0}=70$ $\rm km~s^{-1}$ Mpc$^{-1}$ and a \citet{Kroupa 2001} initial mass function (IMF).

\section{Data}\label{Sec2}

\subsection{The ALFALFA sample}
One of the {\sc Hi} sample used in this work is drawn from the Arecibo Legacy Fast ALFA (ALFALFA) survey \citep{Giovanelli 2005}.  We use the $\alpha.70$ catalog, which includes HI sources exacted from 70\% of the final survey area\footnote{http://egg.astro.cornell.edu/alfalfa/data/index.php}. This catalog contains $\sim$ 25,000 HI sources at $z<0.06$, with more than 95\% having identified optical counterparts. The H{\sc i} mass ($M_{\rm HI}$) was calculated by
\begin{equation}
\frac{M_{\rm HI}}{\rm M_\sun}=(\frac{2.325\times 10^5}{1+z})(\frac{D_{\rm L}^2}{\rm Mpc})(\frac{S_{\rm 21}}{\rm Jy~kms^{-1}}),
\end{equation}\label{eq1}
where $D_{\rm L}$ is the luminosity distance of the source, $z$ is redshift and $S_{\rm 21}$ is the integrated 21-cm line flux density. $D_{\rm L}$ is determined using the Hubble Law for the sources that with $cz>6000~\rm km~s^{-1}$. For sources with $cz<6000~\rm km~s^{-1}$, $D_{\rm L}$ is derived using the local universe peculiar velocity model of \citet{Marsters 2005}. The $\alpha.70$ catalog also provides the coordinates of the matched optical counterparts and the signal-to-noise ratio (S/N) of the \Hi~sources.

The optical data are drawn from the Sloan Digital Sky Survey (SDSS, \citealt{York 2000}). We cross-matched the ALFALFA sources that with confident (or marginal) H{\sc i} detections (with H{\sc i} line detection flag 1 (or 2), see the $\alpha.70$ catalog) with the SDSS data release 7 (DR7) database via the SDSS online SQL tool \footnote{http://skyserver.sdss.org/dr7/en/tools/crossid/crossid.asp}, yielding a sample of 17,134 galaxies.

Of the 17,134 SDSS-ALFALFA matched galaxies, 13,067 ones have spectroscopic redshifts and stellar mass ($M_{\ast}$) estimates \footnote{The $M_{\ast}$ of the spec-z galaxies are available at http://www.mpa-garching.mpg.de/SDSS/DR7/.}. In this paper we wish to estimate the $M_{\ast}$ of the 17,134 galaxies in a uniform way. We first derive the $i$-band absolute magnitude of individual galaxies using the SDSS photometry and the distance information given in the ALFALFA catalog. As confirmed by previous works, the optical broad-band color is in good correlation with the mass-to-light ratio ($M/L$) of galaxies \citep{Bell 2003, Taylor 2011, Fang 2013}. In the left panel of Figure~\ref{fig1}, we show $M/L_{\rm i}$ as a function of $g-i$ color for the 13,067 galaxies that with $M_{\ast}$ estimates. The $g-i$ color has been corrected for Galactic extinction based on the extinction map of \citet{Schlegel 1998}. Since the log$(M/L_{\rm i})$ vs. ($g-i$) relation appears to have a slightly shallower slope at red colors, we fit this relation with a broken linear function:
\begin{eqnarray}
 \mathrm{log}(M/L_{i})=
\begin{cases}
1.0(g-i)+0.93,       & g-i\leq 0.8 \\
\\
0.85(g-i)+1.05,       & g-i\geq 0.8 \\
\end{cases}
\end{eqnarray}\label{eq2}

We compare the stellar masses derived using equation (2) with those of the MPA catalog and show the result in the right panel of Figure~\ref{fig1}. As can be seen, stellar masses derived from equation (2) are well consistent with those from the JHU/MPA catalog. In the following sections, $M_{\ast}$ is referred to which we estimated using equation (2). We further restrict galaxies to have $M_{\ast}>10^{9.0}M_{\sun}$ and $z>0.01$, yielding a final sample of 11,378 galaxies.

\begin{figure*}
\centering
\includegraphics[width=140mm,angle=0]{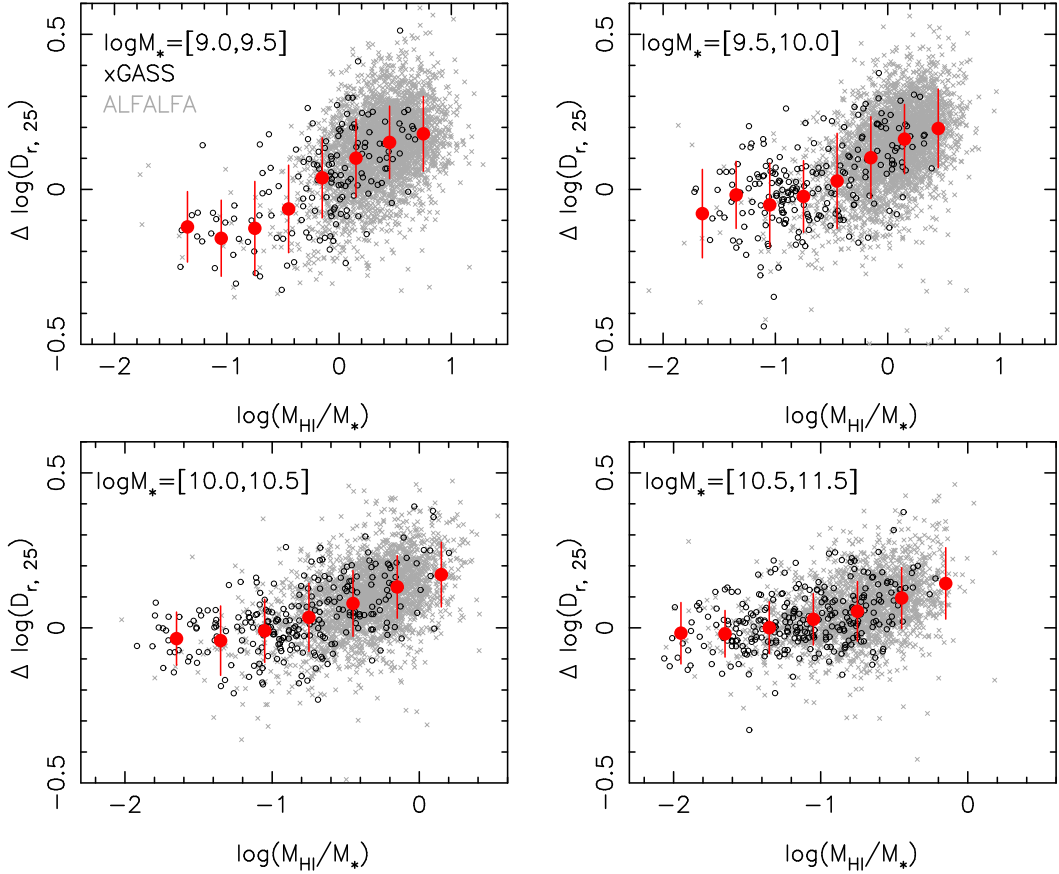}
\caption{Offset from the size$-$mass relation, $\Delta$log$D_{r,~25}$, as a function of \Hi-to-stellar mass ratio ($M_{\rm HI}/M_{\ast}$). In each panel, the large solid circles represent the median $\Delta$log$D_{r,~25}$ in that $M_{\rm HI}/M_{\ast}$ bin, with a bin size of $\Delta$log$(M_{\rm HI}/M_{\ast})=0.3$. Error bars represent the 1$\sigma$ standard deviation.}\label{fig3}
\end{figure*}

\subsection{The xGASS representative sample}
The ALFALFA survey is relatively shallow and biased to \Hi -rich galaxies. To extend our study to more \Hi-deficient galaxies, we supplement the \Hi~sample with galaxies drawn from xGASS (extended GALEX Arecibo SDSS Survey, see \citealt{Catinella 2018}). xGASS is a census of the \Hi~content of 1179 galaxies using the Arecibo telescope, and the galaxies are selected from the intersected area of SDSS, GALEX and ALFALFA. The galaxy sample of xGASS spans a mass range of $M_{\ast}>10^{9.0}M_{\sun}$ at redshift $0.01<z<0.05$. To optimize the survey efficiency, galaxies that with reliable \Hi~detection in ALFALFA were not observed again. The rest galaxies are observed until the \Hi~line is detected, or a limit in $M_{\rm HI}/M_{\ast}$ is reached. The HI spectrum of each galaxy is flagged according to its quality. A catalog containing the \Hi~and other properties (such as star formation rates (SFRs), concentration index, etc.) of the 1179 galaxies is compiled by \citet{Catinella 2018} \footnote{The catalog is available at https://xgass.icrar.org/data.html}. We draw data from the xGASS catalog and convert $M_{\ast}$ from \citet{Chabrier 2003} to \citet{Kroupa 2001} IMF by multiplying a factor of 1.06.  A (1+$z$) factor is multiplied to the xGASS $M_{\rm HI}$ to ensure that both the $M_{\rm HI}$ of ALFALFA and xGASS are calculated with a same method. In the following analysis, we only use the 804 galaxies that with reliable \Hi~detections.

\section{Result}\label{Sec3}
\subsection{Impacts of \Hi~ on the size-mass relation}
In this paper, the size of a galaxy is defined as $D_{\rm r,~25}=2R_{r,~25}$, where $R_{r,~25}$ is the radius at which the SDSS $r-$band surface brightness reaches $\mu_{\rm r}=25.0~\rm mag~arcsec^{-2}$. We use this size indicator because it gives a better description to the global size of a galaxy than the effective radius $R_{\rm e}$, as mentioned above. The quality of SDSS imaging data allows a robust measurement of surface brightness profile of galaxies down to $\mu_{\rm r}=26-27$ $\rm mag~arcsec^{-2}$ \citep{Pohlen 2006, Wang 2018}. In this work, $R_{r,~25}$ is drawn from the SDSS DR7 photometric database. We have checked the quality of the SDSS-measured $R_{r,~25}$ and confirmed that it is suitable for our statistical study (see Appendix A).

In Figure~\ref{fig2}, we show the $D_{\rm r,~25}-M_{\ast}$ relation for the ALFALFA and xGASS galaxy samples. Symbols are color coded by \Hi~gas fraction, defined as $f_{\rm HI}=M_{\rm HI}/M_{\ast}$. Since the $D_{\rm r,~25}-M_{\ast}$ relations of the two samples both show a shallower slope at the low-mass end, we use a second-order polynomial to fit this relation over log$(M_{\ast}/M_{\sun})=[9.0,11.5]$. As shown in the small panels of Figure~\ref{fig2}, the residual of the fitting is independent of stellar mass. The dispersion of the $D_{\rm r,~25}-M_{\ast}$ relation is $\sigma \sim 0.12$ dex, which is significantly smaller than that of the classical $R_{\rm e}-M_{\ast}$ relation \citep{Shen 2003,vanderwel 2014}, as also reported by some previous studies \citep{Cortese 2012,Trujillo 2020}. A most important new feature revealed in Figure~\ref{fig2} is that, at fixed $M_{\ast}$, galaxies that are more rich in \Hi~tend to have larger sizes. This feature also holds when a new size indicator defined by the radial position of a given isomass contour is utilized (see Appendix C).

As shown in previous studies, at fixed $M_{\ast}$, galaxies that with larger $R_{\rm e}$ tend to have bluer colors and younger stellar ages \citep{Shen 2003, Lange 2015, Scott 2017}. Since blue galaxies also tend to be rich in gas, the trend between the sizes and gas fraction of galaxies shown in Figure~\ref{fig2} is expected.

We further divide our samples into 4 stellar mass bins and examine the relation between $\Delta$log$D_{r,~25}$ and $f_{\rm HI}$. $\Delta$log$D_{r,~25}$ is defined as the offset of a galaxy from the size$-$mass relation, i.e., $\Delta$log$D_{\rm r,~25}$=log$D_{\rm r,~25}-$log$\overline{D}_{\rm r,~25}$, where log$\overline{D}_{\rm r,~25}$ is the median size of galaxies as fitted from the xGASS sample. The results are shown in Figure~\ref{fig3}. When replacing log$\overline{D}_{\rm r,~25}$ with that fitted from the ALFALFA sample, the conclusion is unchanged. Overall, the trends between $\Delta$log$D_{r,~25}$ and $f_{\rm HI}$ are similar in all panels. In the high $f_{\rm HI}$ regime, galaxies that are more rich in \Hi~tend to have larger $\Delta$log$D_{r,~25}$. However, when $f_{\rm HI}$ gets sufficiently low, this correlation breaks down, i.e., $\Delta$log$D_{r,~25}$ no longer depends on $f_{\rm HI}$.

The turn over gas fraction ($f_{\rm t}$) at which the $\Delta$log$D_{r,~25}-f_{\rm HI}$ begins to break down seems to depend on $M_{\ast}$. A visual inspection of Figure~\ref{fig3} suggests that at log$(M_{\ast}/M_{\sun})<10.0$, log$f_{\rm t}=[-1.0,-0.7]$, while at log$(M_{\ast}/M_{\sun})>10.0$, log$f_{\rm t}=[-1.5,-1.0]$. We speculate that $f_{\rm t}$ could be somewhat related to the depth of the \Hi~survey used.  Nevertheless, for the mass range we probed, it should be safe to conclude that $\Delta$log$D_{r,~25}$ positively correlates with $f_{\rm HI}$ at log$f_{\rm HI}>-0.7$.

Figure~\ref{fig3} clearly demonstrates that there exists a positive correlation between size and \Hi~fraction at log$f_{\rm HI}>-0.7$. If an isomass-defined size indicator is used, this trend slightly weakens but still exists (see Appendix C). This suggests that the $M/L$ effect also contributes to dispersion of $D_{\rm r,~25}$ at fixed $M_{\ast}$. We do not use an isomass-defined size indicator in the main body of this paper because this will significantly narrow down the dynamical range of size at fixed $M_{\ast}$, making it difficult to examine the trend between size and \Hi~fraction. However, if an isomass-defined size indicator is used, the main conclusions of this work should also hold.

For convenience, we refer galaxies with log$f_{\rm HI}>-0.7$ and log$f_{\rm HI}<-0.7$ as \Hi-rich and \Hi-poor galaxies, respectively. For \Hi-rich galaxies, we interpret the positive correlation between $\Delta$log$D_{r,~25}$  and $f_{\rm HI}$ as a natural consequence of the ``inside-out" disk formation scenario \cite[e.g.,][]{Chiappini 1997,Wang 2011,Perez 2013,Pan 2015}. In theory, normal SFGs are assumed to live in a quasi-equilibrium phase balanced by gas accretion, star formation and outflows. In this case, galaxies with a high $f_{\rm HI}$ preferentially have a higher gas accretion rate \citep{Wang 2013}. In the cold mode accretion, the accreted materials usually have high specific angular momentum that help forming an extended disk component \citep{Stewart 2011,Stewart 2013}, which naturally results in a positive $\Delta$log$D_{r,~25}-f_{\rm HI}$ correlation.

The weak correlation between $\Delta$log$D_{\rm r,~25}$ and $f_{\rm HI}$ exhibited in the \Hi-poor regime implies that \Hi-poor galaxies have a different disk growth mode compared to their gas-rich counterparts. As shown below, \Hi-poor galaxies mainly distribute below the star formation main sequence (SFMS). This suggests that many \Hi-poor galaxies are quenching their star formation, or have been quenched.  During the quenching phase, external cold gas replenishment is expected to be blocked and galaxies have left the star-forming quasi-equilibrium phase \citep{Peng 2015}. In this case, $f_{\rm HI}$ simply reflects the amount of the residual gas reservoirs of galaxies. We thus suggest that the weak correlation between size and \Hi~content shown in the log$f_{\rm HI}<-0.7$ regime is not at odds with the inside-out disk formation scenario.

\begin{figure}
\centering
\includegraphics[width=80mm,angle=0]{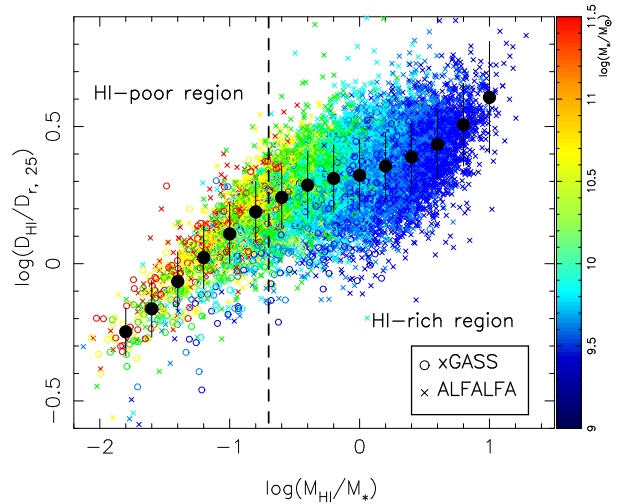}
\caption{The correlation between \Hi-to-stellar size ratio ($D_{\rm HI}/D_{\rm r,~25}$) and \Hi-to-stellar mass ratio ($M_{\rm HI}/M_{\ast}$). Symbols are color coded by $M_{\ast}$. The dashed line marks log$(M_{\rm HI}/M_{\ast})=-0.7$, which separates the parameter plane into an \Hi-poor and an \Hi-rich region. The large solid circles represent the median log$(D_{\rm HI}/D_{\rm r,~25})$ in that $M_{\rm HI}/M_{\ast}$ bin, with a bin size of $\Delta$log$(M_{\rm HI}/M_{\ast})=0.2$. Error bars represent the 1$\sigma$ standard deviation.}\label{fig4}
\end{figure}

\subsection{The \Hi-to-stellar size ratio}
Observationally, a SFG typically has an \Hi~gaseous disk that is more extended than its stellar disk. As stars are formed from cold gas, a comparison between the sizes of \Hi~disk and stellar disk may provide insights in understanding the role of \Hi~in regulating size growth. In the literature, the \Hi~size of a galaxy, $D_{\rm HI}$, is defined as the diameter at which the \Hi~surface density reaches $\Sigma_{\rm HI}=1~\rm M_{\sun}\rm pc^{-2}$. Since both the ALFALFA and xGASS do not provide resolved \Hi~images, in this work we use an indirect method to estimate $D_{\rm HI}$ for individual galaxies. As first reported by \citet{Broeils 1997}, there exists a tight correlation between $D_{\rm HI}$ and $M_{\rm HI}$. With a sample of $\sim 500$ galaxies compiled from literatures, \citet{Wang 2016} revisited the $D_{\rm HI}-M_{\rm HI}$ relation and found
\begin{equation}
{\rm log}~D_{\rm HI}=(0.506\pm0.003){\rm log}~M_{\rm HI}-(3.293\pm0.009),
\end{equation}
which is very close to the one reported by \citet{Broeils 1997}.

With the large galaxy sample in hand, \citet{Wang 2016} further quantified the dispersion of the $D_{\rm HI}-M_{\rm HI}$ relation, finding $1\sigma \sim 0.06$ dex (or 14\%). This scatter is independent of other galactic properties, such as luminosity, $M_{\rm HI}$, H{\sc i}-to-luminosity ratio, etc. Remarkably, galaxies across the Hubble types (from early-type, Sa-Sd, dIrr to even ultra-diffuse galaxies) all follow a same $D_{\rm HI}-M_{\rm HI}$ relation \citep{Wang 2016,Leisman 2017}. Since the galaxy samples of \citet{Wang 2016} and this work cover a similar range in redshift and $M_{\rm HI}$, in this paper we use equation (3) to infer $D_{\rm HI}$ for the ALFALFA and xGASS galaxies. We discuss in more detail in Appendix B to show that this application should be valid.

Figure~\ref{fig4} shows $D_{\rm HI}/D_{\rm r,~25}$ as a function of $f_{\rm HI}$. Inspired by Figure~\ref{fig3}, we first divide the galaxies into 2 subsamples according to $f_{\rm HI}$, with a demarcation line of log$f_{\rm HI}=-0.7$. For the \Hi-poor sample, $D_{\rm HI}/D_{\rm r,~25}$ is quite tightly correlated with $f_{\rm HI}$, in the sense that more \Hi-poor galaxies have a lower $D_{\rm HI}/D_{\rm r,~25}$ ratio. For the \Hi-rich sample, the correlation between these two quantities significantly weakens. Note that there exists an up-bending in the $(D_{\rm HI}/D_{\rm r,~25})-f_{\rm HI}$ relation at the high $f_{\rm HI}$ end. This feature is due to selection effect. We have checked and confirmed that a significant fraction of the galaxies with $M_{\ast}<10^{9.0}M_{\sun}$ have similar $f_{\rm HI}$ but lower $D_{\rm HI}/D_{\rm r,~25}$ compared to their massive counterparts. When galaxies with $M_{\ast}<10^{9.0}M_{\sun}$ galaxies are included, the up-bending feature shown at the high $f_{\rm HI}$ is absent.

At the low-$f_{\rm HI}$ end, Figure~\ref{fig4} should have missed some very low-$f_{\rm HI}$ galaxies due to the \Hi~detection limits of xGASS and ALFALFA. In the log$f_{\rm HI}<-1.0$ regime, the $D_{\rm HI}/D_{\rm r,~25}$ ratio is mainly driven by $D_{\rm HI}$ (or $M_{\rm HI}$) since $D_{\rm r,~25}$ is no longer sensitive to $f_{\rm HI}$ (see Figure~\ref{fig3}). Therefore, the \Hi-undetected galaxies are expected to display in the extrapolation of the sequence defined by the \Hi-detected galaxies with log$f_{\rm HI}<-1.0$.

\begin{figure}
\centering
\includegraphics[width=80mm,angle=0]{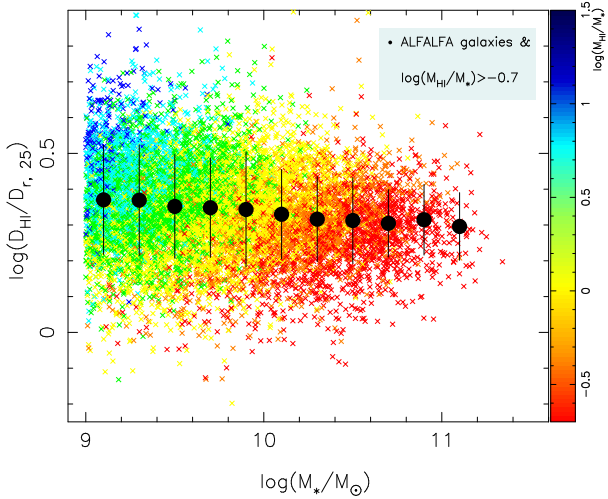}
\caption{The correlation between $D_{\rm HI}/D_{\rm r,~25}$ and $M_{\ast}$ for the \Hi-rich ALFALFA galaxies. Symbols are color coded by $M_{\rm HI}/M_{\ast}$. The large solid circles represent the median log$(D_{\rm HI}/D_{\rm r,~25})$ in that $M_{\ast}$ bin, with a bin size of $\Delta$log$(M_{\ast}/M_{\sun})=0.2$. Error bars represent the 1$\sigma$ standard deviation.}\label{fig5}
\end{figure}

\begin{figure}
\centering
\includegraphics[width=80mm,angle=0]{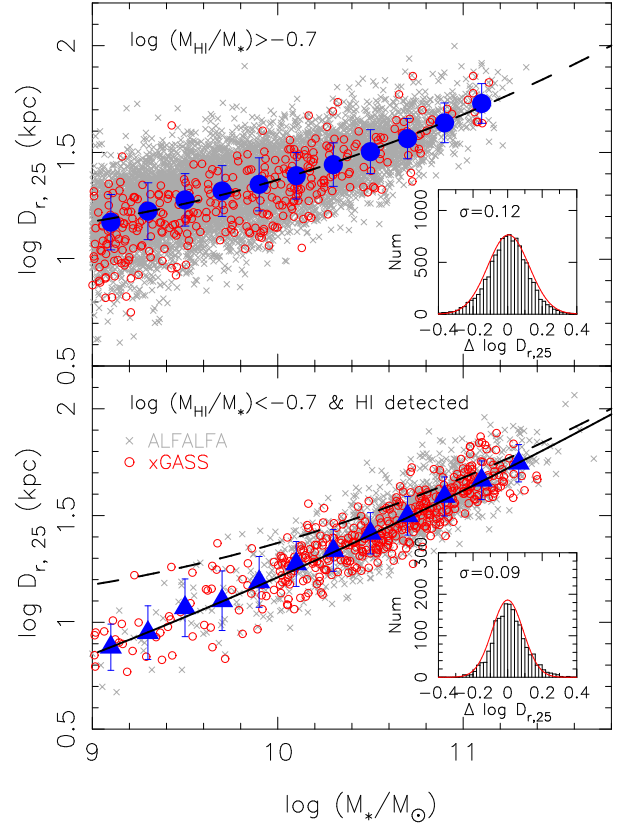}
\caption{Top: the log($D_{\rm r,~25}$)$-$log($M_{\ast}$) relation for the \Hi-rich galaxies.  Large symbols represent the median log $D_{\rm r,~25}$ at each $M_{\ast}$ bin, with a bin size of $\Delta$log$(M_{\ast}/M_{\sun})=0.2$. The dashed line represents the best fit of the log($D_{ \rm r,~25}$)$-$log($M_{\ast}$) relation. The small panel show the size scatter distribution around the best fit relation, and the red line shows the best fit gaussian function. Bottom: the log($D_{\rm r,~25}$)$-$log($M_{\ast}$) relations for the \Hi-poor galaxies, with a best fit shown in solid line. }\label{fig6}
\end{figure}

Since $D_{\rm HI}$ is inferred from $M_{\rm HI}$, Figure~\ref{fig4} thus also shows the interplay between $M_{\ast}$, $D_{\rm r,~25}$ and $M_{\rm HI}$ similar to Figure~\ref{fig2} and Figure~\ref{fig3}. After transforming $M_{\rm HI}$ into $D_{\rm HI}$, one can find that the gas disk shrinks more significantly compared to the stellar disk once galaxies entering the gas-poor phase. This is likely due to the different disk growth behaviors at different galaxy evolutional phases, as we will argue below.

To investigate the \Hi-to-stellar size ratio of \Hi-rich galaxies more specifically, we show $D_{\rm HI}/D_{\rm r,~25}$ as a function of $M_{\ast}$ in Figure~\ref{fig5}. Interestingly, the \Hi-to-stellar size ratio of these galaxies is weakly correlated with $M_{\ast}$. With a sample of 108 galaxies observed by the Westerbork Synthesis Radio Telescope, \citet{Broeils 1997} found that $D_{\rm HI}/D_{\rm r,~25}$ is only weakly correlated with luminosity. We confirm the finding of \citet{Broeils 1997} with a much larger sample and emphasize that this trend only holds for \Hi-rich galaxies.

At this point, it is worthy to revisit the size$-$mass relation separately for the \Hi-rich and \Hi-poor galaxies. The results are shown in Figure~\ref{fig6}. To be consistent with the previous sections, we also use a second-order polynomial to fit the size$-$mass relation.  Overall, the size$-$mass relation of \Hi-rich galaxies has a shallower slope and a larger dispersion than the one of \Hi-poor galaxies. These features resemble those of the size$-$mass relations of star-forming and quiescent galaxies (see Figure 3 of \citealt{vanderwel 2014}). On the one hand, this similarity is straightforward to interpret because the SFRs and \Hi~richness of galaxies are correlated, in the sense that SFGs tend to be more rich in \Hi~compared to quiescent galaxies. On the other hand, an \Hi~view of the size$-$mass relation provides new insights in understanding the origin of size scatter at fixed $M_{\ast}$ (also see Figure~\ref{fig3}). In the next section, we will show that the size scatter of SFGs is correlated with \Hi~richness, rather than SFRs.

\begin{figure}
\centering
\includegraphics[width=80mm,angle=0]{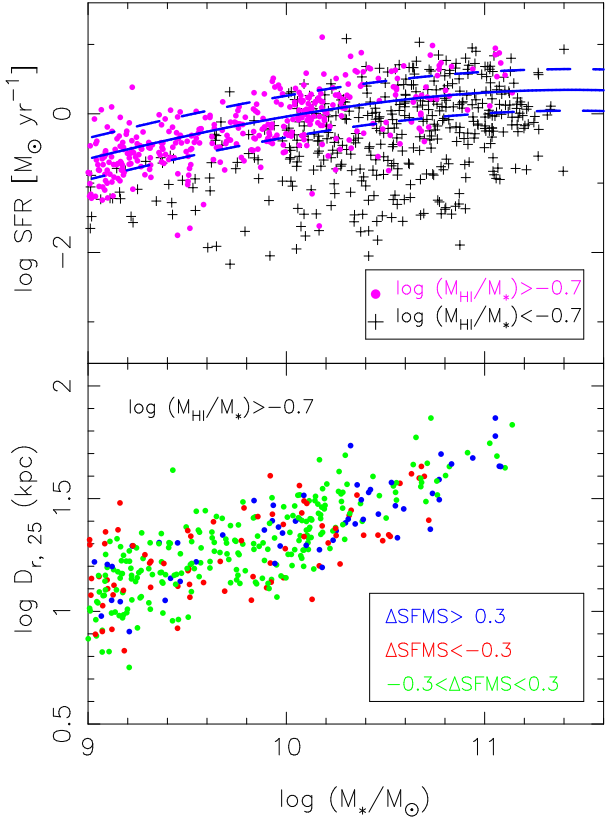}
\caption{Top: the log(SFR)$-$log$M_{\ast}$ relation for xGASS galaxies. \Hi-rich and \Hi-poor galaxies are indicated by solid circles and crosses, respectively. The solid line indicate ridge line of the star formation main sequence defined by \citet{Saintonge 2016}, while the dashed lines indicate $\pm0.3$ dex scatter around the main sequence. Bottom: the log($D_{\rm r,~25}$)$-$log($M_{\ast}$) relation for the \Hi-rich galaxies drawn form the xGASS sample. Galaxies are color coded by their SFR offset from the star formation main sequence ($\Delta SFMS$).}\label{fig7}
\end{figure}
\subsection{connections between Star formation rate, \Hi~fraction and Size}
In a merger-free case, the size of a galaxy grows through in-situ star formation. This naturally predicts a link between size growth and star formation activity. To our knowledge, star formation is more physically correlated with molecular gas, rather than \Hi~gas. Why the size of \Hi-rich galaxies strongly correlate with the \Hi~content? Is the size growth of \Hi-rich galaxies due to enhanced star formation activities?

To answer these questions, we first examine the distribution of xGASS galaxies in the SFR$-M_{\ast}$ plane to explore the correlation between \Hi-richness and SFRs. The SFMS of xGASS galaxies is fitted by \citet{Saintonge 2016}, as shown in the blue solid line of Figure~\ref{fig7}. It can be seen that \Hi-rich galaxies distribute along the SFMS, i.e, they are mostly SFGs. In contrast, \Hi-poor galaxies distribute more broadly, and the majority of them distribute below the SFMS. In low-mass regime of log$(M_{\ast}/M_{\sun})<10.2$, the \Hi-poor galaxies mainly distribute below the SFMS. At the high-mass end, a significant fraction of the \Hi-poor galaxies still distribute on the SFMS. This implies that many massive SFGs may be short in \Hi~gas supply. \citet{Saintonge 2016} also argued that the flattening of the SFMS at the high mass end is primarily due to the decrease of cold gas reservoirs.
\begin{figure}
\centering
\includegraphics[width=80mm,angle=0]{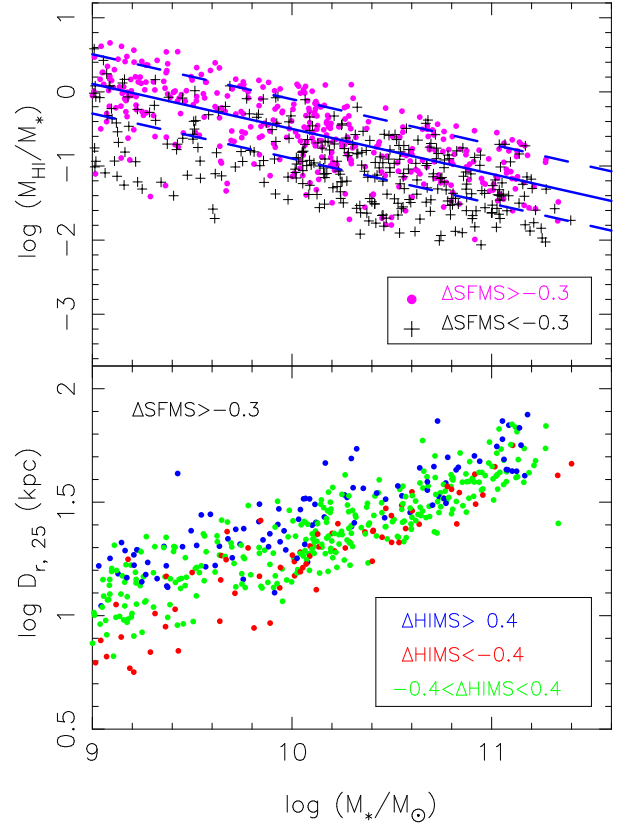}
\caption{Top: the log$(M_{\rm HI}/M_{\ast})-$log$M_{\ast}$ relation for xGASS galaxies. Galaxies with $\Delta SFMS>-0.3$ and $\Delta SFMS<-0.3$ are indicated by pink and black symbols, respectively. The solid line indicate ridge line of the gas main sequence fitted from the galaxies with $\Delta SFMS$>-0.3, while the dashed lines indicate $\pm0.4$ dex scatter around it. Bottom: the log($D_{\rm r,~25}$)$-$log($M_{\ast}$) relation for the $\Delta SFMS>-0.3$ galaxies drawn form the xGASS sample. Galaxies are color coded by their offset from the \Hi~gas main sequence ($\Delta HIMS$). }\label{fig8}
\end{figure}

To explore the correlation between size and SFRs, we divide the \Hi-rich xGASS galaxies into 3 subsamples according to their offset from the SFMS ridge line ($\Delta SFMS$, defined as $\Delta SFMS=\rm log(SFR/SFR_{\rm MS})$) and examine their locations in the size$-$mass plane.  We only include \Hi-rich galaxies since they are likely to live in an actively disk-growing phase, as discussed above. The result is shown in the bottom panel of Figure~\ref{fig7}. Clearly, the location of galaxies in the size$-$mass plane is weakly dependent on $\Delta SFMS$, i.e, there is little correlation between SFR and size at fixed $M_{\ast}$. This result is broadly consistent with that reported in \citet{Lin 2020}.

In the top panel of Figure~\ref{fig8}, we show \Hi~fraction as a function of $M_{\ast}$. Similar to the SFMS, we fit an \Hi~gas main sequence (HIMS) using xGASS galaxies near the SFMS (with $\Delta SFMS$>-0.3), as shown in the blue solid line. It can be seen that galaxies with $\Delta SFMS$>-0.3 generally have higher \Hi~fraction than those with $\Delta SFMS$<-0.3, but these two samples have significantly overlap in the $M_{\rm HI}/M_{\ast}-M_{\ast}$ diagram. \cite{Jano 2020} have investigated the HIMS based on the xGASS sample and reported a similar feature. When investigating the molecular gas fraction vs. stellar mass diagram, \cite{Jano 2020} found that the overlap is significantly reduced (see their Fig.2 and Fig.3).

In the bottom panel of Figure~\ref{fig8}, we examine the size$-$mass relation of SFGs that with $\Delta SFMS>-0.3$. Symbols are color coded by the offset of galaxies from the HIMS.  As can be seen, at fixed $M_{\ast}$, \Hi-rich SFGs tend be more large in size. To conclude, at fixed $M_{\ast}$, the size scatter of SFGs is more physically linked with their \Hi~content, rather than SFRs.

\section{Discussion}\label{Sec3}

\subsection{size growth in the gas-rich phase}
The central finding of this work is that the size of an \Hi-rich galaxy is closely linked to its \Hi~fraction, rather than SFR. Since the majority of star formation occurs inside the stellar disk, a direct interpretation is that \Hi-rich galaxies have a more extended star-forming disk than the \Hi-normal ones. Previous works have studied the extension of star-forming disk through ultraviolet (UV) observations and investigated its relation to the \Hi~properties. \cite{Cortese 2012} showed that the UV/optical size ratio of galaxies is tightly correlated with their \Hi~fraction, in the sense that \Hi-rich galaxies tend to have larger UV/optical size ratios (see their Figure 6). This is qualitatively consistent with our findings.

In an inside-out disk growth scenario, the materials that build a stellar disk is primarily accreted from the surrounding environment. In this sense, the properties of a galaxy would be to a large extent shaped by the properties of the accreted material (such as specific angular momentum, accretion rate, etc). Based on a suite of 30 cosmological magneto-hydrodynamical zoom simulations, \cite{Grand 2017} find that the sizes of galaxies are closely linked to the angular momentum of halo material. Galaxies with the largest disk size are produced by quiescent mergers that deposit high-angular momentum material into the pre-existing disc, simultaneously increasing the spin of dark matter and gas in the halo. In the observational work of \cite{Huang 2012}, the authors studied the spin parameter $\lambda$ for the ALFALFA galaxies, finding that \Hi-rich galaxies tend to reside in high$-\lambda$ halos. A similar conclusion is reached by the recent work of \cite{Mancera 2021}, who show that gas-rich disk galaxies tend to have high specific angular momentum. This can be interpreted since gas with high specific angular momentum is more difficult to collapse and convert into stars, which naturally results in a high $M_{\rm HI}/M_{\ast}$ ratio. We thus suggest that angular momentum is the key parameter behind the relation between size and \Hi~fraction.

Our study also suggests that cold gas accretion should exist in the majority of \Hi-rich galaxies. As shown in Figure~\ref{fig7}, these galaxies are mostly normal SFGs. If gas replenishment does not exist, the \Hi~disks of galaxies will shrink due to continuous gas consumption via star formation. In addition, without the acquisition of angular momentum from accreted materials \citep{Stewart 2011,Stewart 2013}, kinematic evolution will drive the remaining gas settling towards galactic central regions, leading to a shrink of the gaseous disk. As a result, $D_{\rm HI}/D_{\rm r,~25}$ will decrease towards lower $f_{\rm HI}$. Figure~\ref{fig4} demonstrates that this is true at log$f_{\rm HI}<-0.7$, suggesting that \Hi-poor galaxies likely have their \Hi~supply shut down.

In the ``bathtub" model \cite[e.g.,][]{Bouche 2010,Dave 2011, Lilly 2013,Forbes 2014, Peng 2014}, galaxies will approach a quasi-equilibrium  phase when cold gas replenishment, star formation and gas outflows reach a balance. The weak correlation between $D_{\rm HI}/D_{\rm r,~25}$ and $M_{\ast}$ at log$f_{\rm HI}>-0.7$ is likely a manifestation of this phase. There is other observational evidence supporting that galaxies will experience such a phase during their lifetime, for example, the existence of a tight SFMS among SFGs from $z=0$ to $z=6$ \cite[e.g.,][]{Brinchmann 2004, Noeske 2007, Elbaz 2007, Karim 2011,Speagle 2014,Tasca 2015}. Galaxies on the ridge of the SFMS typically have disk-like structures \citep{Wuyts 2011}, suggesting that disk growth is actively occurring in this phase.

\subsection{size growth in the gas-poor phase}

In the \Hi-poor regime, we find that size is weakly correlated with \Hi~fraction. As shown in Figure~\ref{fig7}, galaxies that with log$f_{\rm HI}<-0.7$ typically distribute below the SFMS. Note that a significant fraction of massive SFGs also have log$f_{\rm HI}<-0.7$, and we speculate many of them are undergoing star formation quenching. By modeling the galaxy number density in the NUV$-u$ color space, \citet{Lian 2016} concluded that at log($M_{\ast}/M_{\sun}$)$=$10.5 about 45\% of the SFGs are undergoing quenching. For Milky-Way mass SFGs, \citet{Pan 2017} suggested that their mass budget has been dominated by a quenched component, also implying on-going quenching processes are actively taking place at the massive end. Although many massive galaxies still appear ``star-forming" at present, the quenching progress will push them migrating onto the ``dead sequence" with a time scale of $2-4$ gigayears \citep{Schawinski 2014, Peng 2015, Lian 2016,Hahn 2017}.

In the scenario of star formation quenching, \Hi-poor galaxies grow their sizes primarily via mergers. As shown in \cite{Catinella 2010}, \Hi-poor galaxies preferentially have bulge-dominated (early-type) morphologies. \cite{Shen 2003} show that the size$-$mass relation of SDSS early-type galaxies (ETGs) is consistent with the assumption that they are the remnants of major mergers of present-day disks. In this sense, the size scatter of ETGs should be more physically related with their merger histories, rather than the current gas content.

However, it should be noted that not every \Hi-poor galaxy is necessarily to be quenched or undergoing quenching. For example, numerical studies show that SFGs in the lower envelope of the SFMS typically have lower gas fraction and longer gas depletion time scale ($\tau_{\rm dep}$) than their counterparts on the SFMS ridge \citep{Tacchella 2016}. These galaxies will quench star formation if their gas replenishment time scale ($\tau_{\rm rep}$) is longer than $\tau_{\rm dep}$. In the case of $\tau_{\rm rep}<\tau_{\rm dep}$, they will evolve back to the SFMS ridge and become gas rich again. It is difficult to quantify the fraction of galaxies with $\tau_{\rm rep}<\tau_{\rm dep}$ in the \Hi-poor regime, but the size growth of such galaxies should also follow the scenario discussed for the \Hi-rich galaxies.

\section{Summary}
In this paper, we study the role of \Hi~gas in regulating size growth of local galaxies using galaxy samples from the xGASS and ALFALFA survey. In the \Hi-rich regime, galaxies that are more rich in \Hi~tend to have larger sizes, which we interpret as a natural consequence of the ``inside-out" disk assembly. This trend is absent in the \Hi-poor regime, indicating that size growth is barely affected by \Hi~gas when it has declined to a sufficiently low level. We also study the relations between size, \Hi-fraction and SFR, finding that size is more intrinsically linked with \Hi~fraction, rather than SFR. The \Hi-to-stellar size ratio of \Hi-rich galaxies is found to be weakly dependent on $M_{\ast}$. We conclude that in the \Hi-rich phase, size growth is primarily achieved by star formation. The size dispersion at fixed $M_{\ast}$ is probably driven by the angular momentum of the accreted materials.

\acknowledgments
We are grateful to the anonymous referee for useful suggestions that helped improving the presentation of this paper. This work is supported by the National Key Research and Development Program of China (2017YFA0402703), the National Science Foundation of China (11773076, 12073078), and the Chinese Academy of Sciences (CAS) through a China-Chile Joint Research Fund (CCJRF 1809) administered by the CAS South America Center for Astronomy
(CASSACA).

We wish to thank the entire ALFALFA and xGASS team for providing reduced \Hi~catalog for the community. Funding for SDSS-III has been provided by the Alfred P. Sloan Foundation, the Participating Institutions, the National Science Foundation, and the U.S. Department of Energy Office of Science. The SDSS-III web site is http://www.sdss3.org/

\appendix\label{sec:app}

\section{The reliability of SDSS-measured $R_{25}$ }
It is known that the automatic SDSS photometry pipeline is not reliable for angularly large sources, which is mainly due to the problems in background substraction and blending large galaxies into multiple sources \citep{West 2010}. Since in this work we are going to use $D_{\rm r,~25}=2R_{\rm r,~25}$ as a size indicator, we first need to check whether $R_{\rm r,~25}$ is reliably measured by the SDSS photometry pipeline.

\begin{figure}
\centering
\includegraphics[width=70mm,angle=0]{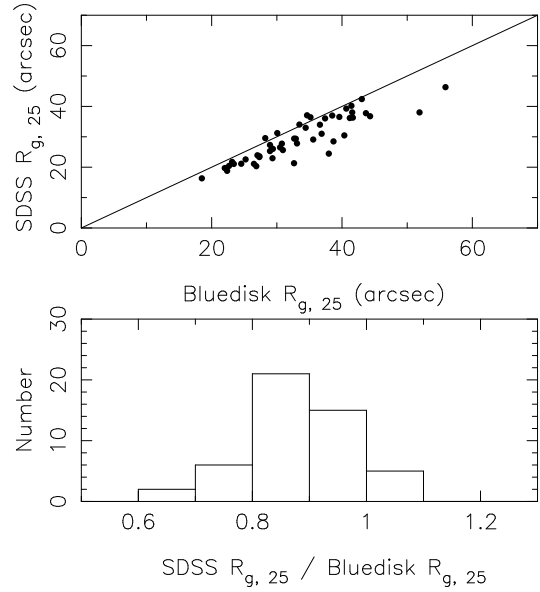}
\caption{Top: The comparison between the $g-$band $R_{\rm 25}$ measured by SDSS and those measured by \cite{Wang 2013} for the Blue disk sample. The solid line indicates 1:1. Bottom: The distribution of the ratio between SDSS $R_{\rm 25}$ and Wang's $R_{\rm 25}$. }\label{fig9}
\end{figure}

To do this, we compare the $g-$band isophotal radii at the surface brightness of 25.0 $\rm mag~arcsec^{-2}$ ($R_{\rm 25}$) provided by the SDSS pipeline with those measured by \cite{Wang 2013} for the Blue disk sample, which contains 50 galaxies with log($M_{\ast}/M_{\sun})=[10.0,11.0]$. The result is shown in Figure~\ref{fig9}. As can be seen, the SDSS-measured $g-$band $R_{\rm 25}$ is slightly smaller than those measured by the Wang's pipeline ($R_{g, 25, SDSS}\approx0.9R_{g, 25, Bluedisk})$. This is due to the systematics of these two pipelines. Since the $g-$band $R_{\rm 25}$ is similar to that measured in the $r-$band, we conclude that the SDSS $R_{\rm 25}$ measurement is reliable at least for galaxies with $R_{\rm 25}<45.0$ arcsec.

\begin{figure}
\centering
\includegraphics[width=70mm,angle=0]{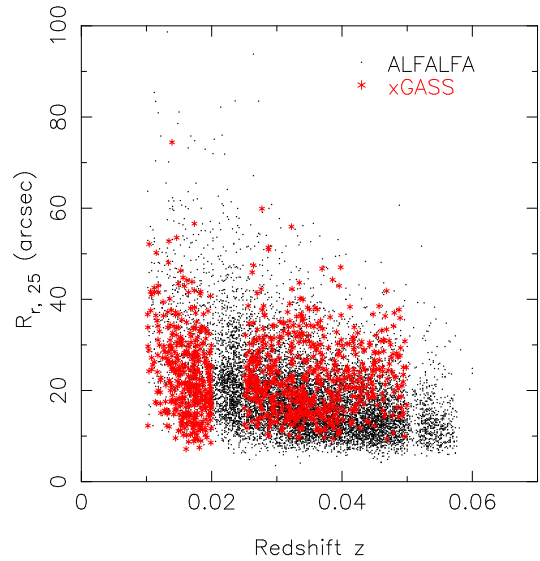}
\caption{The distribution of galaxies in the apparent $R_{\rm r,~25}$ vs. redshift $z$ space. }\label{fig10}
\end{figure}

Figure~\ref{fig10} shows the $r-$band $R_{\rm 25}$ as a function of redshift for the ALFALFA and xGASS samples. As can be seen, the majority of the galaxies have $R_{\rm 25}<45.0$ arcsec, i.e., they are not very extended sources. We thus conclude that the $R_{\rm r,~25}$ provided by the SDSS database can be used in our statistical study.

\section{Do ALFALFA and xGASS galaxies follow the $D_{\rm HI}-M_{\rm HI}$ relation?}
The $D_{\rm HI}$ presented in this work is inferred from the $D_{\rm HI}-M_{\rm HI}$ relation, rather than from direct measurements. To make our analysis work, we first need to verify that the ALFALFA and xGASS \Hi-detected galaxies also follow the $D_{\rm HI}-M_{\rm HI}$ relation. Since spatially-resolved \Hi~imaging data are not available for ALFALFA and xGASS, a directly investigation of the $D_{\rm HI}-M_{\rm HI}$ relation for ALFALFA and xGASS is not feasible. We thus use an indirect approach to access whether the ALFALFA and xGASS galaxies follow the $D_{\rm HI}-M_{\rm HI}$ relation. We do this by placing the ALFALFA and xGASS galaxies together with \citet{Wang 2016} sample in the $M_{\ast}-M_{\rm HI}$ and $M_{\ast}-D_{\rm r,~25}$ plane to examine the parameter distributions of these samples. At fixed $M_{\ast}$, if ALFALFA and xGASS \Hi-detected galaxies were more \Hi-deficient than Wang's sample, then it may be problematic to apply the $D_{\rm HI}-M_{\rm HI}$ relation to ALFALFA and xGASS. If not, the ALFALFA and xGASS galaxies should also follow the the same $D_{\rm HI}-M_{\rm HI}$ relation established by \citet{Wang 2016}.

\begin{figure*}
\centering
\includegraphics[width=140mm,angle=0]{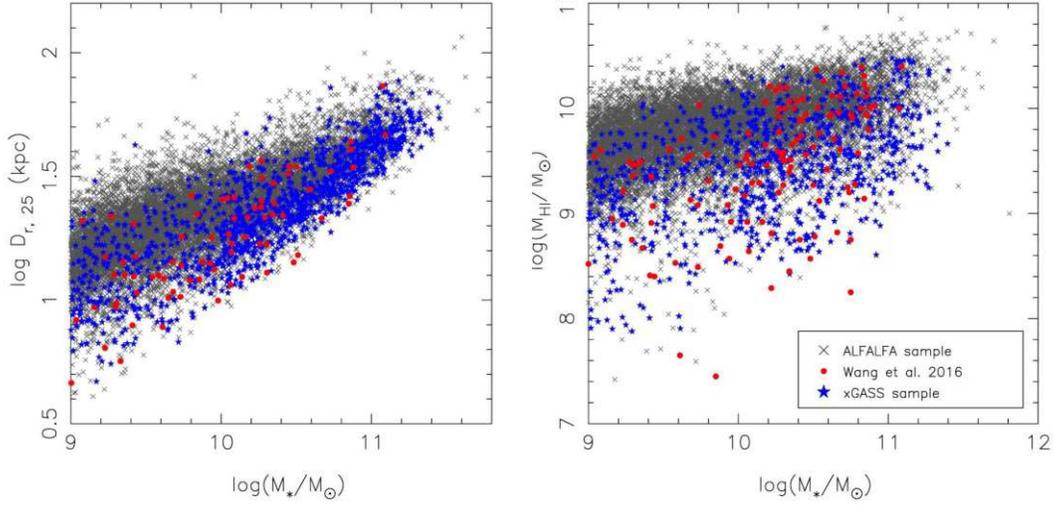}
\caption{Left: the distribution of ALFALFA, xGASS and \citet{Wang 2016}'s sample in the $M_{\ast}-D_{\rm r,~25}$ plane. Right: the $M_{\ast}-M_{\rm HI}$ plane. At a given stellar mass, it is clear that ALFALFA detected galaxies appear more gas rich than Wang's sample. At log$(M_{\ast}/M_{\sun})>9.0$, both xGASS and Wang's sample cover a similar parameter region. Given this, the application of the $D_{\rm HI}-M_{\ast}$ relation of \citet{Wang 2016} to ALFALFA and xGASS should be valid.}\label{fig11}
\end{figure*}

To derive the stellar mass for Wang's sample, we cross-identified the sample galaxies with SDSS footprints, yielding a sample of 195 galaxies. We then estimated the stellar masses of these galaxies using equation (2). In left and right panel of Figure~\ref{fig11}, we show the parameter distribution of these three samples in the $M_{\ast}-D_{\rm r,~25}$ and $M_{\ast}-M_{\rm HI}$ space, respectively. Overall, the xGASS galaxies cover a similar region as Wang's sample in both parameter spaces. It can be seen that at log$(M_{\ast}/M_{\sun})=9.0-10.0$, the ALFALFA galaxies appear to be systematically more gas rich than Wang's sample. This is because the ALFALFA survey is biased to \Hi~rich galaxies. At high-masses, these two samples cover a similar parameter region, which is due to the fact that Wang's sample includes 23 galaxies that with unusually high \Hi~mass fraction drawn from the Bluedisk project \citep{Wang 2013}.  Since Wang's sample well follow the $D_{\rm HI}-M_{\rm HI}$ relation, we are thus confident that the \Hi-detected galaxies from ALFALFA and xGASS should also follow this relation since they do not appear to be more \Hi-deficient.

\begin{figure*}
\centering
\includegraphics[width=140mm,angle=0]{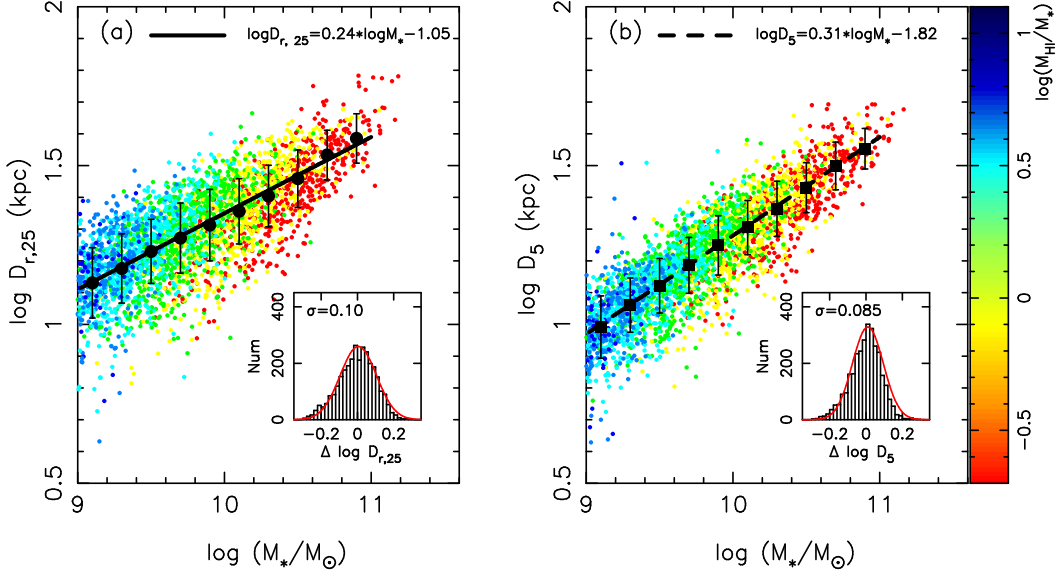}
\caption{Left:the log($D_{\rm r,~25}$)$-$log($M_{\ast}$) relation for face-on \Hi-rich galaxies. The large symbols represent the median log$(D_{\rm r,~25})$ in that $M_{\ast}$ bin, with a bin size of $\Delta$log$(M_{\ast}/M_{\sun})=0.2$. Error bars represent the 1$\sigma$ standard deviation. We use a linear function to fit this relation, and the best-fit result is shown in the solid line. Small symbols are color coded by log$(M_{\rm HI}/M_{\ast})$. The small panel shows the size scatter around the best fit relation. Right: the log($D_{\rm 5}$)$-$log($M_{\ast}$) relation. }\label{fig12}
\end{figure*}

\section{Replacing $D_{\rm r, ~25}$ with $D_{\rm 5}$}
In this work, the size of a galaxy is defined by $D_{\rm r,~25}$. Taking the variation of mass-to-light ratio ($M/L$) among galaxies into account, a size definition with fixed stellar mass surface density could be more physically meaningful \citep{Trujillo 2020}. We have selected a sample of $\sim$3700 ALFALFA galaxies with log$f_{\rm HI}>-0.7$ to compute their stellar mass density profiles based on the SDSS public photometric data. The sample galaxies are selected to have minor-to-major axis ratio of $b/a>0.7$, $R_{\rm 25}>10\farcs0$ and $0.01<z<0.06$.  The SDSS pipeline measures the surface brightness profile of galaxies in five bands ($ugriz$) in a series of circular annuli of fixed angular size. For each galaxy, we first corrected the annular photometry for Galactic extinction and $k-$correction for each annulus. Then a spline was fit to the cumulative light profile (to conserve flux), which was then differentiated to derive the surface brightness profile interpolated over a grid of angular sizes in $0\farcs1$. The surface brightness profile was then converted into stellar mass surface density profile using the mass-to-light ratio expressed in equation (2).

We select a new sample rather than use the original sample presented in the main body of the paper to do this test because: (1) only at log$f_{\rm HI}>-0.7$ galaxy size shows a clear dependence on \Hi~richness; (2) it is difficult to measure the stellar mass density profile for edge-on galaxies. For this sample, we use $D_{\rm 5}$ to define the size of a galaxy, where $D_{\rm 5}$ is the diameter at which the stellar mass surface density reaches $5~ M_{\sun}\rm pc^{-2}$. In \cite{Trujillo 2020}, the authors use $D_{\rm 1}$ as a size indicator. We do not use $D_{\rm 1}$ because our photometric data is shallower than those used in \cite{Trujillo 2020}, which may result in large uncertainties in the $D_{\rm 1}$ measurement. In left and right panel of Figure~\ref{fig12}, we show the $M_{\ast}-D_{\rm r,~25}$ and $M_{\ast}-D_{\rm 5}$ relations for our sample galaxies, respectively. To be consistent with \cite{Trujillo 2020}, we use a linear function to fit both relations. As can be seen, the $M_{\ast}-D_{\rm 5}$ relation is tighter and steeper than the $M_{\ast}-D_{\rm r,~25}$ relation. A similar feature is also reported by \cite{Trujillo 2020}. At fixed stellar mass, \Hi-rich galaxies also tend to have larger $D_{\rm 5}$, although this feature is less evident than that revealed in the $M_{\ast}-D_{\rm r,~25}$ relation. We thus conclude that the positive correlation between size and \Hi~richness for \Hi-rich galaxies can not be fully due to the $M/L$ effect.


\begin{thebibliography}{99}

\bibitem[Bell et al.(2003)]{Bell 2003} Bell, E.~F., McIntosh, D.~H., Katz, N., \& Weinberg, M.~D.\ 2003, \apjs, 149, 289
\bibitem[Bouch{\'e} et al.(2010)]{Bouche 2010} Bouch{\'e}, N., Dekel, A., Genzel, R., et al.\ 2010, \apj, 718, 1001
\bibitem[Broeils \& Rhee(1997)]{Broeils 1997} Broeils, A.~H., \& Rhee, M.-H.\ 1997, \aap, 324, 877
\bibitem[Brown et al.(2015)]{Brown 2015} Brown, T., Catinella, B., Cortese, L., et al.\ 2015, \mnras, 452, 2479
\bibitem[Brinchmann et al.(2004)]{Brinchmann 2004} Brinchmann, J.,
Charlot, S., White, S.~D.~M., et al.\ 2004, \mnras, 351, 1151
\bibitem[Catinella et al.(2010)]{Catinella 2010} Catinella, B., Schiminovich, D., Kauffmann, G., et al.\ 2010, \mnras, 403, 683
\bibitem[Catinella et al.(2018)]{Catinella 2018} Catinella, B., Saintonge, A., Janowiecki, S., et al.\ 2018, \mnras, 476, 875
\bibitem[Chabrier(2003)]{Chabrier 2003} Chabrier, G.\ 2003, \pasp, 115, 763
\bibitem[Chiappini et al.(1997)]{Chiappini 1997} Chiappini, C., Matteucci, F., \& Gratton, R.\ 1997, \apj, 477, 765
\bibitem[Chen et al.(2020)]{Chen 2020} Chen, X., Wang, J., Kong, X., et al.\ 2020, \mnras, 492, 2393
\bibitem[Cortese et al.(2012)]{Cortese 2012} Cortese, L., Boissier, S., Boselli, A., et al.\ 2012, \aap, 544, A101
\bibitem[Dale et al.(2016)]{Dale 2016} Dale, D.~A., Beltz-Mohrmann, G.~D., Egan, A.~A., et al.\ 2016, \aj, 151, 4.
\bibitem[Dav{\'e} et al.(2011)]{Dave 2011} Dav{\'e}, R., Finlator, K., \& Oppenheimer, B.~D.\ 2011, \mnras, 416, 1354
\bibitem[de Vaucouleurs(1948)]{de Vaucouleurs 1948} de Vaucouleurs, G.\ 1948, Annales d'Astrophysique, 11, 247
\bibitem[Elbaz et al.(2007)]{Elbaz 2007} Elbaz, D., Daddi, E., Le Borgne, D., et al.\ 2007, \aap, 468, 33
\bibitem[Fang et al.(2013)]{Fang 2013} Fang, J.~J., Faber, S.~M., Koo, D.~C., \& Dekel, A.\ 2013, \apj, 776, 63
\bibitem[Frankel et al.(2019)]{Frankel 2019} Frankel, N., Sanders, J., Rix, H.-W., et al.\ 2019, \apj, 884, 99
\bibitem[Forbes et al.(2014)]{Forbes 2014} Forbes, J.~C., Krumholz, M.~R., Burkert, A., \& Dekel, A.\ 2014, \mnras, 438, 1552
\bibitem[Giovanelli et al.(2005)]{Giovanelli 2005} Giovanelli, R., Haynes, M.~P., Kent, B.~R., et al.\ 2005, \aj, 130, 2598
\bibitem[Graham(2019)]{Graham 2019} Graham, A.~W.\ 2019, \pasa, 36, e035
\bibitem[Grand et al.(2017)]{Grand 2017} Grand, R.~J.~J., G{\'o}mez, F.~A., Marinacci, F., et al.\ 2017, \mnras, 467, 179
\bibitem[Hahn et al.(2017)]{Hahn 2017} Hahn, C., Tinker, J.~L., \& Wetzel, A.\ 2017, \apj, 841, 6
\bibitem[Huang et al.(2012)]{Huang 2012} Huang, S., Haynes, M.~P., Giovanelli, R., \& Brinchmann, J.\ 2012, \apj, 756, 113
\bibitem[Huang et al.(2014)]{Huang 2014} Huang, S., Haynes, M.~P., Giovanelli, R., et al.\ 2014, \apj, 793, 40
\bibitem[Hopkins et al.(2009)]{Hopkins 2009} Hopkins, P.~F., Hernquist, L., Cox, T.~J., et al.\ 2009, \apj, 691, 1424
\bibitem[Janowiecki et al.(2020)]{Jano 2020} Janowiecki, S., Catinella, B., Cortese, L., et al.\ 2020, \mnras, 493, 1982
\bibitem[Karim et al.(2011)]{Karim 2011} Karim, A., Schinnerer,
E., Mart{\'{\i}}nez-Sansigre, A., et al.\ 2011, \apj, 730, 61
\bibitem[Kauffmann(2015)]{Kauffmann 2015} Kauffmann, G.\ 2015, \mnras, 450, 618
\bibitem[Yim \& van der Hulst(2016)]{Yim 2016} Yim, K. \& van der Hulst, J.~M.\ 2016, \mnras, 463, 2092
\bibitem[Kroupa(2001)]{Kroupa 2001} Kroupa, P.\ 2001, \mnras, 322,
231
\bibitem[Lange et al.(2015)]{Lange 2015} Lange, R., Driver, S.~P., Robotham, A.~S.~G., et al.\ 2015, \mnras, 447, 2603
\bibitem[Leisman et al.(2017)]{Leisman 2017} Leisman, L., Haynes, M.~P., Janowiecki, S., et al.\ 2017, \apj, 842, 133
\bibitem[Lian et al.(2016)]{Lian 2016} Lian, J., Yan, R., Zhang, K., \& Kong, X.\ 2016, \apj, 832, 29
\bibitem[Lilly et al.(2013)]{Lilly 2013} Lilly, S.~J., Carollo, C.~M., Pipino, A., Renzini, A., \& Peng, Y.\ 2013, \apj, 772, 119
\bibitem[Lin et al.(2020)]{Lin 2020} Lin, L., Faber, S.~M., Koo, D.~C., et al.\ 2020, \apj, 899, 93
\bibitem[Mancera Pi{\~n}a et al.(2021)]{Mancera 2021} Mancera Pi{\~n}a, P.~E., Posti, L., Pezzulli, G., et al.\ 2021, arXiv:2107.02809
\bibitem[Marsters (2005)]{Marsters 2005} Masters,K. L., 2005, PhD thesis, Cornell Univ.

\bibitem[Noeske et al.(2007)]{Noeske 2007} Noeske, K.~G., Weiner, B.~J., Faber, S.~M., et al.\ 2007, \apjl, 660, L43
\bibitem[Pan et al.(2015)]{Pan 2015} Pan, Z., Li, J., Lin, W., et al.\ 2015, \apjl, 804, L42
\bibitem[Pan et al.(2017)]{Pan 2017} Pan, Z., Zheng, X., \& Kong, X.\ 2017, \apj, 834, 39
\bibitem[Peng \& Maiolino(2014)]{Peng 2014} Peng, Y.-j., \& Maiolino, R.\ 2014, \mnras, 443, 3643
\bibitem[Peng et al.(2015)]{Peng 2015} Peng, Y., Maiolino, R., \& Cochrane, R.\ 2015, \nat, 521, 192
\bibitem[P{\'e}rez et al.(2013)]{Perez 2013} P{\'e}rez, E., Cid Fernandes, R., Gonz{\'a}lez Delgado, R.~M., et al.\ 2013, \apjl, 764, L1
\bibitem[Pichon et al.(2011)]{Pichon 2011} Pichon, C., Pogosyan, D., Kimm, T., et al.\ 2011, \mnras, 418, 2493
\bibitem[Pohlen \& Trujillo(2006)]{Pohlen 2006} Pohlen, M. \& Trujillo, I.\ 2006, \aap, 454, 759
\bibitem[Saintonge et al.(2016)]{Saintonge 2016} Saintonge, A., Catinella, B., Cortese, L., et al.\ 2016, \mnras, 462, 1749.
\bibitem[Schawinski et al.(2014)]{Schawinski 2014} Schawinski, K., Urry, C.~M., Simmons, B.~D., et al.\ 2014, \mnras, 440, 889
\bibitem[Scott et al.(2017)]{Scott 2017} Scott, N., Brough, S., Croom, S.~M., et al.\ 2017, \mnras, 472, 2833
\bibitem[Schlegel et al.(1998)]{Schlegel 1998} Schlegel, D.~J., Finkbeiner, D.~P., \& Davis, M.\ 1998, \apj, 500, 525
\bibitem[Shen et al.(2003)]{Shen 2003} Shen, S., Mo, H.~J., White, S.~D.~M., et al.\ 2003, \mnras, 343, 978
\bibitem[Speagle et al.(2014)]{Speagle 2014} Speagle, J.~S., Steinhardt, C.~L., Capak, P.~L., \& Silverman, J.~D.\ 2014, \apjs, 214, 15
\bibitem[Stewart et al.(2011)]{Stewart 2011} Stewart, K.~R., Kaufmann, T., Bullock, J.~S., et al.\ 2011, \apj, 738, 39
\bibitem[Stewart et al.(2013)]{Stewart 2013} Stewart, K.~R., Brooks, A.~M., Bullock, J.~S., et al.\ 2013, \apj, 769, 74
\bibitem[Taylor et al.(2011)]{Taylor 2011} Taylor, E.~N., Hopkins, A.~M., Baldry, I.~K., et al.\ 2011, \mnras, 418, 1587
\bibitem[Tacchella et al.(2016)]{Tacchella 2016} Tacchella, S., Dekel, A., Carollo, C.~M., et al.\ 2016, \mnras, 457, 2790.
\bibitem[Tasca et al.(2015)]{Tasca 2015} Tasca, L.~A.~M., Le F{\`e}vre, O., Hathi, N.~P., et al.\ 2015, \aap, 581, A54
\bibitem[Trujillo et al.(2020)]{Trujillo 2020} Trujillo, I., Chamba, N., \& Knapen, J.~H.\ 2020, \mnras, 493, 87
\bibitem[van der Wel et al.(2014)]{vanderwel 2014} van der Wel, A., Franx, M., van Dokkum, P.~G., et al.\ 2014, \apj, 788, 28
\bibitem[Wang et al.(2011)]{Wang 2011} Wang, J., Kauffmann, G., Overzier, R., et al.\ 2011, \mnras, 412, 1081
\bibitem[Wang et al.(2013)]{Wang 2013} Wang, J., Kauffmann, G., J{\'o}zsa, G.~I.~G., et al.\ 2013, \mnras, 433, 270
\bibitem[Wang et al.(2016)]{Wang 2016} Wang, J., Koribalski, B.~S., Serra, P., et al.\ 2016, \mnras, 460, 2143
\bibitem[Wang et al.(2018)]{Wang 2018} Wang, J., Zheng, Z., D'Souza, R., et al.\ 2018, \mnras, 479, 4292
\bibitem[West et al.(2010)]{West 2010} West, A.~A., Garcia-Appadoo, D.~A., Dalcanton, J.~J., et al.\ 2010, \aj, 139, 315
\bibitem[Wuyts et al.(2011)]{Wuyts 2011} Wuyts, S., F{\"o}rster Schreiber, N.~M., van der Wel, A., et al.\ 2011, \apj, 742, 96
\bibitem[Y{\i}ld{\i}z et al.(2017)]{Yildiz 2017} Y{\i}ld{\i}z, M.~K., Serra, P., Peletier, R.~F., Oosterloo, T.~A., \& Duc, P.-A.\ 2017, \mnras, 464, 329
\bibitem[York et al.(2000)]{York 2000} York, D.~G., Adelman, J., Anderson, J.~E., Jr., et al.\ 2000, \aj, 120, 1579
\end{thebibliography}
\end{document}